\documentclass[twocolumn,showpacs,preprintnumbers,amsmath,amssymb,floatfix,nofootinbib]{revtex4-1}
\usepackage{graphicx}
\usepackage{dcolumn}
\usepackage{bm}
\usepackage{amsthm}
\usepackage{amsmath}
\usepackage{amsfonts}
\usepackage{color}

\usepackage{caption}
\usepackage{subcaption}

\newcommand{\ket}[1]{\left| #1 \right>} 
\newcommand{\bra}[1]{\left< #1 \right|} 

\newcommand{\be}{\begin{equation}}
\newcommand{\ee}{\end{equation}}
\newcommand{\ba}{\begin{eqnarray}}
\newcommand{\ea}{\end{eqnarray}}

\newcommand{\n}[1]{\label{#1}}
\newcommand{\eq}[1]{(\ref{#1})}

\begin{document}

\title{Velocity Effects on an Accelerated Unruh-DeWitt Detector}
\author{Shohreh Abdolrahimi}
\email{abdolra@ualberta.ca}
\affiliation{Institut f\"ur Physik, Universit\"at Oldenburg, Postfach 2503 D-26111 Oldenburg, 
Germany}
\begin{abstract}
We analyze the response of an Unruh-DeWitt detector moving along an unbounded spatial trajectory in a two-dimensional spatial plane with constant independent magnitudes of both the four-acceleration and of a timelike proper time derivative of the four-accelration. In a Fermi-Walker frame moving with the detector, the direction of the acceleration rotates at a constant rate around a great circle. This is the motion of a charge in a uniform electric field when in the frame of the charge there is both an electric and a magnetic field. We compare the response of this detector to a detector moving with constant velocity in a thermal bath of the corresponding temperature for non-relativistic velocities, and two regimes ultraviolet and infrared. In infrared regime, the detector in the Minkowski space-time moving along the spatially two-dimensional trajectory should move with a higher speed to keep up with the same excitation rate of the inertial detector in a thermal bath. In ultraviolet regime, the dominant modification in the response of this detector compared to the black body spectrum of Unruh radiation is the same as the dominant modification perceived by a detector moving with constant velocity in a thermal bath.
\end{abstract}

\maketitle

\section{Introduction}
A uniformly accelerated observer in Minkowski spacetime, i.e. linearly accelerated observer with constant proper
acceleration, associates a thermal bath of Rindler particles to the no-particle state of inertial observers. This is the {\it Unruh effect} \cite{Un, Fulling, Da}. It implies the conceptually important result that the particle content
of a field theory is observer dependent. The Unruh effect is important in its own right, perhaps having experimental applications in particle accelerators \cite{Bell1, Bell2,Unruh2,Unruh3}, electrons in Penning traps \cite{Rogers,Brown}, atoms in microwave cavities \cite{Scully,Belyanin}, or hadronic collisions \cite{Barshay1,Barshay2,Kharzeev}, and as a
tool to investigate other phenomena such as the thermal
emission of particles from black holes \cite{Haw1,Haw2} and cosmological horizons \cite{Gibbons}. For a review of the Unruh effect and applications see \cite{Rev1}. Recently,  Mart\'in-Mart\'inez, Fuentes, and Mann have shown, \cite{Martin}, that a detector acquires a Berry phase due to its motion in spacetime and this fact can be used for the direct detection of the Unruh effect in regimes physically accessible with current technology.

Unruh has introduced a detector model consisting of a
small box containing a non-relativistic particle satisfying
the Schr\"odinger equation \cite{Un}. The system is said to have detected
a quanta if the particle in the box jumps from
the ground state to some excited state. DeWitt \cite{De} has introduced a detector which consists of a two-level point
monopole. In this paper we use an idealized point detector with internal energy levels labelled by energy $E_{0}$ and $E>E_{0}$, coupled via a monopole interaction with a scalar field $\varphi$, known 
as an Unruh-DeWitt detector in the literature.

An eternal uniformly accelerated Unruh-DeWitt detector, moving along a linear spatial trajectory in Minkowski vacuum with constant magnitude of its four-acceleration $a$, perceives a radiation rate which is equivalent to a detector at rest in a thermal bath of Minkowski particles of the temperature $T=\hbar a/(2\pi c k_{B})$, where  $\hbar$ is the reduced Planck constant, $c$ is the speed of light, and $k_{B}$ is the Boltzmann constant. This detector will experience a time-independent situation and hence settle in a stationary state. However, there exist other time-like curves such that the geodesic interval between two points along the curve depend only on the proper time interval. There are called stationary curves, a classification of such curves into six categories has been done by Letaw \cite{Letaw}. The vacuum excitation spectra of detectors on a representative sample
of such stationary world lines have been calculated, some of which were presented only numerically. The corresponding vacuum states have also been classified. It was shown by Letaw and Pfautsch \cite{Letaw2} that the corresponding vacuum states are found to be restricted to two possibilities: Those in coordinate systems without event horizons are the Minkowski vacuum; those in coordinate systems with event
horizons are the Fulling vacuum. The analog of Unruh effect for spatially circular trajectories has been discussed in particular with relation to polarization effects of electrons in storage rings and for electrons circulating in a cavity \cite{Bell, Bell2, Unruh2,Leinaas, Rogers, Levin, Davis,Akhmedov}. 
Gutti, Kulkarni, and Sriramkumar
have shown that the response of the rotating detector can be computed exactly (albeit, numerically) even when it
is coupled to a field that is governed by a nonlinear dispersion relation \cite{Gutti}.  Korsbakkena, and Leinaasa \cite{Korsbakkena} related the excitation spectrum of a detector moving along planar stationary trajectories to the properties of Minkowski vacuum in the accelerated frame and defined an effective temperature in terms of the transition rate of a detector into up or down states. Barbado and Visser analyzed the response function of an Unruh-DeWitt detector moving with time-dependent acceleration along a one-dimensional trajectory in Minkowski spacetime \cite{Barbado}.   

In this paper, we consider a special case of a stationary trajectory. We consider a detector moving along an unbounded spatial trajectory in a two-dimensional spatial plane with constant independent magnitudes of both the four-acceleration and of a timelike proper time derivative of four-accelration, such that in a Fermi-Walker frame moving with the detector, the direction of the acceleration rotates at a constant rate around a great circle. This is the motion of a charge in a uniform electric field when in the frame of the charge there is both an electric and a magnetic field. We choose a special coordinate for describing the motion, in which one of the components of the 4-velocity, $w=dy/d\tau$ ($\tau$ is the proper-time of the detector), is constant. 
We calculate explicitly the response of an Unruh-DeWitt detector moving along the above trajectory in non-relativistic limit\footnote{When the acceleration is set to zero this detector corresponds to one moving with constant velocity in Minkowski vacuum; such a detector perceives no temperature.}. In non-relativistic limit, the zero order term is of course the thermal spectrum of Unruh radiation, but we are interested to find the next dominant term in the response of the detector, proportional to the square of the four-velocity component  $w^{2}$. On the other hand, we consider a detector moving with constant non-relativistic speed $\tilde{w}$ in a thermal bath of temperature corresponding to the Unruh temperature $T=\hbar a/(2\pi c k_{B})$. The first dominant term in the response of this detector is the Plankian spectrum of a thermal bath, we find the next dominant term in the response of this detector, proportional to $\tilde{w}^{2}$, and compare this dominant term to the one we find from the accelerating detector moving in Minkowski space-time in two regimes, which we call ultraviolet and infrared. 

We begin this paper with a review of the definitions
of the physical quantities involved in the description of an
Unruh-DeWitt detector, Sec I. In Section II, we describe a trajectory of the detector and calculate the response of an Unruh-DeWitt detector following the described trajectory. We compare the response of this Unruh-DeWitt detector with that of a uniformly accelerated detector ($w=0$), moving along spatially straight line, and also with a detector moving with constant velocity in a thermal bath. 
In this paper, we use the system of units where $\hbar=c=k_{B}=1$.

\section{The Detector}
Suppose we have a pointlike two-level system (detector) moving along a worldline described by the functions $x^{\mu}(\tau)=(t(\tau),\bf{x}(\tau))$, where $\tau$ is the detector's proper time, and $\mu$ labels the coordinates in the space-time. Assume that this two-level system has two internal energy levels labelled by the energy $E_{0}$ and  $E>E_{0}$ and is coupled to a quantum scalar field $\varphi$ via a monopole interaction, 
$V=m~q(\tau)\varphi[x(\tau)]$, where $q(\tau)$ is the monopole
moment operator \cite{Un,De,Birrell,Frolov}, and $m$ is the interaction constant. Then, the system, i.e. the two-level detector, and the quantum field is described by the following Hamiltonian
\be
\hat H=\hat H_{0}^{(o)}+\hat H_{0}^{(f)}+\hat{V}, 
\ee
where $\hat H_{0}^{(o)}$ is the Hamiltonian of the free two-level system,
$\hat{H}^{(f)}$ is the Hamiltonian of the free quantum scalar field, and $\hat{V}$ defines the interaction. Assume that $\ket{A}$'s are the eigenvectors of the orthonormal basis of the Hilbert space of the states of the system without interaction,
\be
\hat{H}^{(of)}_{0}\ket{A}=E_{A}^{(of)}\ket{A},~~~~ \hat{H}_{0}^{(of)}=\hat H_{0}^{(o)}+\hat H_{0}^{(f)}.
\ee
For a general trajectory, the system of the two-level detector and the field will not always remain in its ground state $E_{0}^{(of)}$, but will undergo a transition to an excited state $E^{(of)}>E_{0}^{(of)}$. 
If we assume that the interaction constant $m$ is small, in the first-order approximation of the perturbation theory the probability amplitude of the transition from the initial state $\ket{A}$ to the final state $\ket{B}$ at the proper time $\tau$ is given by
\ba
&&\mathcal{A}_{BA}=-i ~m\int_{-\infty}^{\tau} d\tau' V_{BA}(\tau'),\n{pro}\\
&&V_{BA}=\bra{B} \text{exp}({i\hat{H}_{0}^{(of)}\tau}) ~ \hat{\varphi}(0)\hat{q}(0)~ \text{exp}({-i\hat{H}_{0}^{(of)}\tau})\ket{A}.\nonumber\\ \n{pro2}
\ea
Let the states $\ket{n}$ and $\ket{N}$ to be the eigenstates of the non-interacting free Hamiltonian of the two level detector and the non-interacting free Hamiltonian of the free field,
\be
\hat H_{0}^{(o)}\ket{n}=E_{n}\ket{n}, ~~~~~ \hat H_{0}^{(f)}\ket{N}=\omega_{N}\ket{N},
\ee
respectively. Then, the states $\ket{A}$ of the free system of the two-level detector and field can be written as
\be\n{ba}
\ket{A}=\ket{n}\ket{N}.
\ee
The two-level detector is either in the ground state $|0>$ with energy $E_{0}$ or in the excited state $\ket{1}$ with energy $E$. The probability amplitude of the transition \eq{pro} can be derived using \eq{pro2} which in the basis \eq{ba} can be written as
\ba
&&V_{BA}=V_{Mm~Nn}=q_{mn}~ \text{e}^{i(E_{m}-E_{n})\tau} \bra{M}\hat{\varphi}[x(\tau)]\ket{N},\nonumber\\
\ea
where $q_{mn}= \bra{m}q(0)\ket{n}$. Suppose that the field $\varphi$ is initially in vacuum state $\ket{0_{M}}$, where the subscript $M$ stands for Minkowski vacuum, and the two-level system is in ground state $E_{0}$. Let us consider mental copies of the above two-level Unruh-DeWitt detector, where these copies are different only in one sense, the value of their second energy level $E$ is different. Assume that all of these detectors are prepared in the same initial state  and following identical trajectories (see \cite{Barbado} for a discussion about physical construction of such a system of detectors). The transition probability to all possible $\ket{M}$ and $\ket{1}$'s (of different value of energy $E$) for this ensemble of detectors is  
\ba
&&p_{M1~00}=m^{2} \sum_{E}|q_{10}|^{2} \int_{-\infty}^{\tau}d\tau' \int_{-\infty}^{\tau}d\tau'' \text{e}^{i(E-E_{0})\Delta\tau} \nonumber\\
&&\hspace{4cm}\times G^{+}(x(\tau'),x(\tau'')),\n{Int}\nonumber\\
\ea
where $G^{+}$ is the positive frequency Wightman function, 
\be
G^{+}(x(\tau'),x(\tau''))=\bra{0}\hat{\varphi}[x(\tau')]\hat{\varphi}[x(\tau'')]\ket{0},
\ee
which for massless scalar field reads
\be
 G^{+}(x(\tau'),x(\tau''))=-\frac{1}{4\pi^{2}[(t'-t''-i\epsilon)^{2}-|{\bf x'}-{\bf x''}|^{2}]} ~.\n{GW1}
 \ee
 
Here, $\epsilon\ll 0$. Note that $(t',{\bf x'})$ and $(t'',{\bf x''})$ are functions of the proper time. If $G^{+}(x(\tau'),x(\tau''))$ can be written as $G^{+}(\Delta\tau,r)$, where $r=|{\bf x'}-{\bf x''}|$ and $\Delta\tau=\tau'-\tau''$, the integrand in \eq{Int} depends only on $\Delta\tau$, and we can write \eq{Int} in the following form
 \ba 
&&p_{M1~00}=m^{2}  \sum_{E} |q_{10}|^{2} \int_{-\infty}^{\tau}d\tau' \int_{-\infty}^{\infty}d(\Delta\tau) \text{e}^{i(E-E_{0})\Delta\tau}\times\nonumber\\
&&\hspace{3cm}G^{+}(\Delta\tau,r).\nonumber\\
\ea
The transition probability per unit proper time is
\be
p_{\Delta \tau}=\frac{dp_{M1~00}}{d\tau}=m^{2} \sum_{E} |q_{10}|^{2} \mathcal{F}(E),\n{prob1}
\ee
where 
\be
\mathcal{F}(E)=\int_{-\infty}^{\infty}d(\Delta\tau) \text{e}^{i(E-E_{0})\Delta\tau}G^{+}(\Delta\tau,r),\n {Re1}
\ee
is the response function per unit proper time, and is independent of the detailed structure of the detector. If the quantum scalar field is initially in the thermal state rather than the Minkowski vacuum state then the response function $\mathcal{F}$ has to be replaced by
\be
\mathcal{F}_{\beta}(E)=\int_{-\infty}^{\infty}d(\Delta\tau) \text{e}^{i(E-E_{0})\Delta\tau}G^{+}_{\beta}(\Delta\tau,r),\n{InR}
\ee
where $G^{+}_{\beta}$ is the Wightman thermal Green function, which for the case of massless scalar field is \cite{Frolov}
\ba
G^{+}_{\beta}(\Delta\tau,r)&&=G^{+}(\Delta\tau,r)+\frac{1}{4\pi^{2}(\Delta t^{2}-r^{2})},\nonumber\\
&&+\frac{\coth[\pi (r+\Delta t)/\beta]+\coth[\pi (r-\Delta t)/\beta]}{8\pi\beta r},\nonumber\\
\n{GWTe}
\ea
where $\beta=1/T$ is the inverse temperature, and $\Delta t=t'-t''$. In what follows we shall consider a two-level Unruh-DeWitt detector as described in this section. The response function of this detector per unit proper time can be calculated using \eq{Re1} if the detector is moving in Minkowski vacuum or using \eq{InR} if the detector is coupled to the thermal quantum scalar field. 
\section{Motion of the detector}
Consider an Unruh-DeWitt detector explained in the previous section, moving in Minkowski space-time along an unbounded spatial trajectory in a two-dimensional spatial plane with  constant square of magnitude of four-acceleration $a_{\mu}a^{\mu}=a^{2}$, where $a^{\mu}=d^{2}x^{\mu}/d\tau^{2}$, and constant magnitude of a timelike proper-time derivative of four-acceleration $(da_{\mu}/d\tau)( da^{\mu}/d\tau)$, and having component of the four-velocity, ${dy}/{d\tau}=w$, a constant, namely a detector moving along the following worldline 
\ba
&&x^{\mu}(\tau)=\biggl(\frac{a}{\alpha^{2}}\sinh({\alpha \tau}),\frac{a}{\alpha^{2}}\cosh({\alpha \tau}),w\tau,0\biggl),\n{traj}\\
&& \alpha=\frac{a}{\sqrt{1+w^{2}}}>0,\n{traj0}
\ea
where $x^{\mu}=(t,x,y,z)$ are the Minkowski coordinates. This is the motion of a charge in a uniform electric field when in the frame of the charge there is both an electric and a magnetic field. We have chosen a special coordinate for describing the motion, in which one of the components of the 4-velocity, $w = dy/d\tau$ is constant. 
The magnitude of the Fermi-Walker derivative of the acceleration is 
\be
|Da|=\biggl(D_{\mu}^{(F)}[a] D^{\mu~(F)}[a]\biggl)^{\frac{1}{2}}=\frac{a^{2}w}{\sqrt{1+w^{2}}},
\ee
The parameter $\eta = |Da|/a^2$ is less than one. In a Fermi-Walker frame moving with the detector, the direction of the acceleration rotates at a constant rate $\eta$ around a great circle. 
For a circular trajectory rather than \eq{traj}, which gives $\eta>1$, one needs to replace the uniform electric field with a uniform magnetic field and take the charge moving so that in the frame of the charge there is both a magnetic and an electric field. 

For $w=0$, the trajectory is 
\be
x^{\mu}=x^{\mu}(\tau)=(a^{-1}\sinh({a \tau}),a^{-1}\cosh({a\tau}),0,0),\n{trajectory1}
\ee
which is a trajectory of a detector moving along a spatially straight line  along the $x$ direction with constant magnitude of four-acceleration $a_{\mu}a^{\mu}=a^{2}$.

Note that if instead of the trajectory \eq{traj}, with component of the four-velocity $w=dy/d\tau=const.$, we have considered the component of the three-velocity $dy/dt=v$ to be constant, the response of the detector would have been completely equivalent to that of a detector which is moving along a spatially straight line with constant magnitude of  four-acceleration, \eq{trajectory1}, as such observers can be related to the ones moving along the trajectory \eq{trajectory1} by Lorentz transformations.  
\subsection{Motion of the detector in the Minkowski vacuum}
For the trajectory \eq{traj}, the positive frequency Wightman Green function \eq{GW1} for a massless scalar field reads
\be
G^{+}(\Delta\tau)=-\frac{\alpha^{4}}{16\pi^{2}a^{2}}\biggl[\sinh^{2}(\frac{\alpha\Delta\tau}{2}-\frac{i\epsilon \alpha^{2}}{a})-\frac{w^{2}\alpha^{4}}{4a^{2}}\Delta\tau^{2}\biggl]^{-1},\n{GWt}
\ee
Here, we have absorbed a positive function of $\tau$ and $\tau'$, $[\sinh(\alpha\tau)-\sinh(\alpha\tau')]/[\sinh({\Delta\tau}/{2\alpha})\cosh({\Delta\tau}/{2\alpha})]$, into $\epsilon$.
Note that for $w=0$ we have $\alpha=a$ and \eq{GWt} immediately converts to the positive Wightman Green function of a detector moving along a spatially straight line in the $x$ direction \eq{trajectory1}, with constant magnitude of the four-acceleration (spatially one-dimensional)
\be
G^{+ ~(1d)}(\Delta \tau)=-\frac{a^{2}}{16\pi^{2}}\biggl[\sinh^{2}(\frac{a\Delta\tau}{2}-i\epsilon a)\biggl ]^{{-1}}.
\ee
Here, and in what follows the $(1d)$ index is used two distinguish the quantities such as the Wightman Green function, or response function calculated for the spatially one-dimensional trajectory \eq{trajectory1} as opposed to the index $(2d)$ for the quantities associated with the spatially two-dimensional trajectory \eq{traj}. 
The transition probability per unit proper time \eq{prob1} for the detector following trajectory \eq{trajectory1} is 
\ba
&&p_{\Delta \tau}=m^{2} \sum_{E}  |q_{10}|^{2}\mathcal{F}^{(1d)}(E),\\
&&\mathcal{F}^{(1d)}(E)=\frac{\Delta E}{2\pi [\text{e}^{2\pi\Delta E/a}-1]},\n{1dcase}
\ea
where $\Delta E=E-E_{0}$. This is the usual black body excitation rate, indicating that the excitation rate of an accelerated detector coupled to the field $\varphi$ in the state $\ket{0_{M}}$ is the same as that of a detector, unaccelerated, at rest in a bath of thermal radiation at temperature $T={1}/{\beta}=a/(2\pi)$. 

In the ``infrared limit'' $\Delta E\ll 1$, the  black body excitation rate \eq{1dcase} has the following dominant behavior:
\be
\mathcal{F}^{(1d)}(E)\sim \frac{1}{2\pi \beta}.
\ee
In the ``ultraviolet limit'' $\Delta E\gg 1$, the  black body excitation rate \eq{1dcase} has the following dominant behavior:
\be
\mathcal{F}^{(1d)}(E)\sim \frac{\Delta E}{2\pi} \text{e}^{-\beta \Delta E}.
\ee
We calculate the response function per unit proper time \eq{prob1} of the detector following the trajectory \eq{traj} for non-relativistic velocities $v_{y}\ll 1$
\be
v_{y}=\frac{dy}{dt}=\frac{w}{\sqrt{1+w^{2}}\cosh(\alpha \tau)},
\ee
or $w\ll1$.
Note that here we are not considering the ultra-relativistic limit because for ultra-relativistic velocities $v_{y}\rightarrow 1$ or $w\rightarrow\infty$ the response function of the detector \eq{traj} is suppressed, namely the Unruh-effect is suppressed as
\be
\mathcal{F}^{(2d)}(E)\sim\frac{\Delta E}{2\pi [\text{e}^{2\pi\Delta E/a}-1]}\frac{1}{w^{4}}.
\ee
The Wightman Green function \eq{GWt} in the leading order for $w\ll1$ reads
\ba
G^{+(2d)}(\Delta\tau)&&=(1-2w^{2})G^{+ ~(1d)}(\Delta\tau)\nonumber\\
&&\hspace{-1cm}-\frac{a^{2}}{16\pi^{2}}\biggl[\sinh(a\Delta\tau-2i\epsilon a)
(\frac{a\Delta \tau}{4}-i\epsilon a)+\frac{a^{2}\Delta \tau^{2}}{4}\biggl]\times\nonumber\\
&&\biggl[\sinh^{4}(\frac{a\Delta\tau}{2}-i\epsilon a)\biggl ]^{{-1}} w^{2}+\mathcal{O}(w^{4}).\nonumber\\
\ea
To calculate the response function per unit proper time \eq{prob1}, we use the following identity:
\be
\sinh x=x \prod_{k=1}^{\infty}(1+\frac{x^{2}}{k^{2}\pi^{2}}).
\ee
Calculating the integral \eq{Re1}, we arrive to
\be
\mathcal{F}^{(2d)}(E)=\mathcal{F}^{(1d)}({E})+\mathcal{F}^{a}({E})w^{2}+\mathcal{O}(w^{4}),\n{2d}
\ee
where
\ba
&&\mathcal{F}^{a}(E)=-\frac{\text{e}^{\beta \Delta E}\Delta E^{2}}{12\pi\beta[\text{e}^{\beta \Delta E}-1]^{2}}\times\nonumber\\
&\biggl[&\frac{8\pi^{2}}{\Delta E^{2}}+9\beta^{2 } -\beta \Delta E(\frac{4\pi^{2}}{\Delta E^{2}}+{\beta^{2}})\frac{e^{\beta \Delta E}+1}{e^{\beta \Delta E}-1}\biggl]\n{case1}.
\ea
For the ``infrared'' tail of the spectrum $\Delta E\ll 1$, the response function is 
\be
\mathcal{F}^{(2d)}(E)= \frac{1}{2\pi \beta}\biggl[1-(\frac{7}{6}-\frac{\pi^{2}}{9})w^{2}\biggl]+\mathcal{O}(\Delta E).\n{Faappro}
\ee
For the ``ultraviolet'' tail of the spectrum $\Delta E\gg 1$, the excitation rate \eq{case1} has the following dominant behavior:
\be
\mathcal{F}^{(a)}(E)\sim \frac{\Delta E^{3}}{12\pi} \text{e}^{-\beta \Delta E}\beta^{2}. \n{smallA}
\ee
\subsection{Motion of the detector in the thermal bath}
We now consider a detector moving along spatially straight line with constant component of its  four-velocity $\tilde{w}$
\be
x^{\mu}(\tau)=(\sqrt{1+\tilde{w}^{2}}\tau,0,\tilde{w}\tau,0),\n{traj3}
\ee
in a thermal bath of temperature ${T}$, (see \cite{Costa1,Costa2}), corresponding to the temperature that an accelerated Unruh-DeWitt detector moving along a spatially straight line with constant magnitude of four-acceleration, trajectory \eq{trajectory1}, in Minkowski vacuum perceives, i.e. $T=a/(2\pi)$. Here and in what follows by a thermal bath we mean thermal quantum scalar field. We are interested to see if there is any relation between $\mathcal{F}^{(2d)}(E)$, \eq{2d}, and the response function of the detector moving along a spatial line with constant non-relativistic speed $v=\tilde{w}/\sqrt{1+\tilde{w}^{2}}\ll 1$ or $w\ll 1$ in a thermal bath of corresponding temperature ${T}$. We consider a detector with the same parameters as that of previous subsection. From \eq{GWTe} the Wightman thermal Green function for a detector following trajectory \eq{trajectory1} is  
\ba
G^{+}_{\beta}(\Delta\tau)&=&-\frac{1}{4\pi^{2}(\Delta \tau-i\epsilon)^{2}}+\frac{\sqrt{1-v^{2}}}{8\pi \beta v\Delta\tau}\biggl[\coth(\frac{\pi \gamma_{+}\Delta\tau}{\beta})\nonumber\\
&&+\coth(\frac{\pi \gamma_{-}\Delta\tau}{\beta})\biggl]+\frac{1}{4\pi^{2}\Delta \tau^{2}},
\ea
where $\gamma_{\pm}=\sqrt{(1\pm v)/(1\mp v)}$, and $\beta=1/T$. The response function per unit time of this detector is
\be
\mathcal{F}^{th}({E})=\frac{\sqrt{1-v^{2}}}{4\pi \beta v}\ln\biggl[\frac{(1-\text{e}^{-\beta\Delta{E}\gamma_{-}})}{(1-\text{e}^{-\beta\Delta{E}\gamma_{+}})}\biggl],\n{Rtb}
\ee
where $\Delta{E}={E}-{E}_{0}$ is the difference between the ground state and excited state of the detector.

For non-relativistic velocities $w\ll1$ the leading behavior of the response function 
\eq{Rtb} is 
\ba
\mathcal{F}^{th}({E})=\mathcal{F}^{(1d)}({E})+\mathcal{F}^{v}({E})\tilde{w}^{2}+\mathcal{O}(\tilde{w}^{4}),\n{Ftha1}
\ea
where
\ba
\mathcal{F}^{v}({E})=-\frac{\text{e}^{\beta \Delta {E}}\Delta E^{2}}{12\pi\beta[\text{e}^{\beta \Delta {E}}-1]^{2}}
\biggl[3\beta^{2}-\beta^{3} \Delta E\biggl(\frac{e^{\beta \Delta {E}}+1}{e^{\beta \Delta{E}}-1}\biggl)\biggl],\n{case2}\nonumber\\
\ea
and $\mathcal{F}^{(1d)}({E})$ is the same as \eq{1dcase}.

For the infrared tail of the spectrum $\Delta{E}\ll 1$, the $\mathcal{F}^{th}$,\eq{Ftha1}, is \footnote{It is easy to understand \eq{Fthappro}. Consider two observers immersed in the blackbody
radiation, observer $O$ at rest relative to the radiation, thus he/she
sees strictly isotropic blackbody radiation, and the other observer, $O'$ is
moving with speed $v$ along the x-axis of the first observer. The moving
observer carries with him a detector with collecting
area $A$, with its normal at angle $\theta$ to the axis. It has been shown, \cite{11,12,13}, that the Lorentz transformation change the radiation
temperature $T$ to an effective directional radiation temperature $T'$ 
\be
T'(T,v,\theta)=\frac{T\sqrt{1-v^{2}}}{1-v\cos\theta},\n{Ts}
\ee
but the observer $O'$ looking in the
fixed direction $\theta$ still would map out a blackbody
spectrum. Even though the validity of this conclusion has been questioned in \cite{Costa1}, the author has  considered this conclusion to be valid for infrared sector of radiation $\Delta{E}\ll T$. An Unruh-DeWitt detector perceives only the radiation over the whole angles. Integrating \eq{Ts} over the solid angle, the average temperature perceived by an Unruh-DeWitt detector is Lorentz transformed according to $T'=T(1-v^{2}/6)$. In other words $\beta'=\beta(1+v^{2}/6)$ or $\beta'=\beta[1+\tilde{w}/(6\sqrt{1-\tilde{w}^{2}})]$. Plug this transformation of $\beta'$ into the black body excitation rate, and consider the non-relativistic limit $\tilde{w}\ll 1$ and then the infrared limit $\Delta{E}\ll T$, expression \eq{Fthappro} will be reproduced.}
\be
\mathcal{F}^{th}({E})= \frac{1}{2\pi \beta}[1-\frac{1}{6}\tilde{w}^{2}]+\mathcal{O}(\Delta{E}).\n{Fthappro}
\ee
The expression \eq{Fthappro} can be mapped to the expression \eq{Faappro} if we relate the speed of the detector moving in a thermal bath to the one following the trajectory \eq{traj} in Minkowski space by the following transformations 
\be
\tilde{w}=w\sqrt{21-2\pi^{2}}/\sqrt{3}=0.65w.\n{mapping1}
\ee
The detector in the Minkowski space-time moving along trajectory \eq{traj} should move with a higher speed $w=1.54\tilde{w}$ to keep up with the same excitation rate of the inertial detector in a thermal bath   in the infrared limit. 
 
For the ultraviolet tail of the spectrum $E \gg 1$, the excitation rate \eq{case2} has the following dominant behavior:
\be
\mathcal{F}^{(v)}(E)\sim \frac{\Delta E^{3}}{12\pi} \text{e}^{-\beta \Delta E}\beta^{2}. \n{smallA2}
\ee
This is the same as \eq{smallA}. Therefore, in ultraviolet regime, the dominant modification in the response of the detector following trajectory \eq{traj} compared to the black body spectrum of Unruh radiation is the same as the dominant modification perceived by a detector moving with constant four-velocity component $\tilde{w}$ in a thermal bath along the trajectory \eq{traj3}. 
\section{Conclusion}
We have considered the response of an Unruh-DeWitt detector moving along an unbounded spatial trajectory in a two-dimensional spatial plane with constant independent magnitudes of both the four-acceleration $a$ and of a timelike proper time derivative of 
four-accelration, and having component of four-velocity $w=dy/d\tau$ constant. This is the motion of a charge in a uniform electric field when in the frame of the charge there is both an electric and a magnetic field. We have compared the response function of this detector \eq{2d} and \eq{case1} to that of a detector moving with constant velocity in a thermal bath of the corresponding temperature $T=a/(2\pi)$ in non-relativistic limit, in the ultraviolet and in the infrared limit. The dominant term in the response function is the Plank distribution, equivalent to a detector moving along a spatially straight line with constant magnitude of four-acceleration $a$. The second dominant term $\mathcal{F}^{a}$ in the response function of this detector, can be mapped to the second dominant term $\mathcal{F}^{v}$ of a detector moving non-relativisticly in a thermal bath via \eq{mapping1} in the infrared limit. In order to map the response functions of these two detectors in these different situations, the detector in the Minkowski space-time moving along trajectory \eq{traj} should move with a higher speed $w=1.54\tilde{w}$ to keep up with the same excitation rate of the inertial detector in a thermal bath  in the infrared limit.  We also have shown that in ultraviolet regime the dominant modification in the response of the detector following trajectory \eq{traj} compared to the black body spectrum of Unruh radiation is the same as the dominant modification perceived by a detector moving with constant four-velocity component $\tilde{w}$ in a thermal bath along the trajectory \eq{traj3}. 
\begin{acknowledgments}
The author gratefully acknowledges support by the DFG
Research Training Group 1620 ``Models of Gravity''. I would like to thank professor Don N. Page and Dr. Andrey A. Shoom, and  Christos Tzounis for valuable suggestions. I would also like to thank professor Valeri P. Frolov for pointing out the reference \cite{Korsbakkena} to me.

\end{acknowledgments}

\end{document}